\documentclass[fleqn,twoside]{article}
\usepackage{espcrc2}

\usepackage{amsmath}
\usepackage{amssymb}
\usepackage{multicol}
\usepackage{graphics}
\usepackage{epsfig}
\usepackage{latexsym}
\usepackage{nopageno}
\usepackage{pifont}
\usepackage{epic}
\usepackage{eepic}
\usepackage{epsfig}
\usepackage{pstricks}

\newcommand{\mb}{\mathbf}
\newcommand{\beqn}{\begin{eqnarray}}
\newcommand{\eeqn}{\end{eqnarray}}
\newcommand{\beq}{\begin{displaymath}}
\newcommand{\eeq}{\end{displaymath}}
\newcommand{\cttilde}{\tilde{c}_t}

\begin{document}
\title{Properties of the renormalized quark mass in the 
Schr\"{o}dinger functional\\
with a non--vanishing background field}

\author{Stefan Kurth\address[HU]{Institut f\"{u}r Physik, Humboldt-Universit\"{a}t
zu Berlin,\\ Invalidenstr.~110, 10115 Berlin, Germany}} 

\begin{abstract}
We compute the current quark mass in the Schr\"{o}dinger functional with
a non--vanishing background field at one loop order of perturbation theory.
The results are used to obtain the critical mass at which the renormalized
quark mass vanishes, and some lattice artefacts at one loop order.
\end{abstract}

\maketitle

\section{THE SCHR\"{O}DINGER FUNCTIONAL}
The Schr\"odinger Functional method treats QCD on a space-time 
cylinder $L^3\times T$.\\
\unitlength = 0.2cm
\begin{picture}(42,16)
  \thinlines
  \filltype{shade}
  \put(9,7){\ellipse{3}{8}}
  \put(2,12){\makebox(8,2)[r]{$\exp(aC_k)$}}
  \put(25,12){\makebox(8,2)[r]{$\exp(aC_k')$}}
  \filltype{shade}
  \put(25,7){\ellipse{3}{8}}
  \put(9,3){\line(1,0){16}}
  \put(9,11){\line(1,0){16}}
  \put(20,7){\ellipse{3.375}{9}}
  \put(18.3124,7){\line(-1,1){0.5}}
  \put(18.3125,7){\line(1,1){0.5}}
  \put(14.5,7){\makebox(3,1)[r]{$L^3$}}
  \put(7,0){\makebox(4,2)[t]{$x_0=0$}}
  \put(23,0){\makebox(4,2)[t]{$x_0=T$}}
\end{picture}

For the gauge fields, we choose periodic boundary conditions in spatial
direction and 
  Dirichlet boundary conditions
  \beqn
     U(x,k)|_{x_0=0}  &=& \exp\{aC_k\}  \\
     U(x,k)|_{x_0=T}  &=& \exp\{aC'_k\} 
  \eeqn
in time direction.
Here, $C_k$ and $C'_k$ are constant and diagonal, 
imposing a constant
colour electric background field $V(x,\mu)$.

The quark fields have Dirichlet boundary conditions in time direction:
  \beqn
     P_{+}\psi(x)|_{x_0=0} &=& \rho(\mb{x}) \\
     P_{-}\psi(x)|_{x_0=T} &=& \rho'(\mb{x}) \\
     \bar{\psi}(x)P_{-}|_{x_0=0} &=&\bar{\rho}(\mb{x}) \\
     \bar{\psi}(x)P_{+}|_{x_0=T} &=& \bar{\rho}'(\mb{x})
  \eeqn
with the projectors
\begin{equation}
P_{\pm} = \frac{1}{2}(1\pm\gamma_0),
\end{equation}
and are periodic in space: $\psi(x+\hat{k}L) = e^{i\theta}\psi(x)$,
  $\bar{\psi}(x+\hat{k}L) = e^{-i\theta}\bar{\psi}(x)$.

The observables are
  gauge invariant combinations of the fields inside the cylinder,\\
   $\psi(x), \bar{\psi}(x), U(x,\mu)$, \\
  and of the ``boundary quark fields'' 
\beqn
\zeta(\mb{x})=\frac{\delta}{\delta\bar{\rho}(\mb{x})}&,&\quad
 \zeta'(\mb{x})=\frac{\delta}{\delta\bar{\rho}'(\mb{x})},\\
\bar{\zeta}(\mb{x})=-\frac{\delta}{\delta\rho(\mb{x})}&,&\quad
 \bar{\zeta}'(\mb{x})=-\frac{\delta}{\delta\rho'(\mb{x})}.
\eeqn
\section{THE CURRENT QUARK MASS}
\subsection{Definition of the current quark mass}
For the quark mass, we adopt 
the definition of \cite{PCAC} based on the PCAC relation.
For this purpose, we define the bare correlation functions
\beqn
& f_A(x_0) &= -\frac{a^9}{L^3}
\sum_{\mathbf{x},\mathbf{y},\mathbf{z}}\nonumber\\
& & \frac{1}{3}\left\langle A_0^a(x)\bar{\zeta}(\mathbf{y})\gamma_5\frac{1}{2}
\tau^a\zeta(\mathbf{z})\right\rangle, \\
& f_P(x_0) &= -\frac{a^9}{L^3}
\sum_{\mathbf{x},\mathbf{y},\mathbf{z}}\nonumber\\
& &\frac{1}{3}\left\langle P^a(x)\bar{\zeta}(\mathbf{y})\gamma_5\frac{1}{2}
\tau^a\zeta(\mathbf{z})\right\rangle,
\eeqn
  where $A^a$ and $P^a$ denote the axial current and density.

Now we can define the quark mass
\beqn
& &\!\!\!\!\!\!\!\!\! m(x_0) = \nonumber\\
& &\!\!\!\!\!\frac{\frac{1}{2}(\partial_0^* + \partial_0)f_A(x_0)
 + c_Aa\partial_0^*\partial_0f_P(x_0)}{2f_P(x_0)},
\eeqn
which depends on $x_0$ due to cutoff effects.
Here, $c_A$ is the improvement coefficient which is needed to cancel
the ${\rm O}(a)$ discretization error
of the axial current.\\
  As an unrenormalized quark mass, we take $m$ in the centre of the lattice
\begin{equation}
  m_1 = \left\{ \begin{array}{l}
      m\left(\frac T 2\right) \quad\mbox{\black for even $T/a$,} \\
      \frac 1 2 \left( m\left(\frac{T-a}{2}\right)
                      +m\left(\frac{T+a}{2}\right)\right)\\
      \qquad\qquad\mbox{\black for odd $T/a$.}
      \end{array} \right.
\end{equation}

Analogously, a mass $m'$ can be computed using the boundary quark fields
$\zeta'$ and $\bar{\zeta'}$ at $x_0=T$ instead of $\zeta$ and $\bar{\zeta}$.
An alternative mass $m_2$ may be defined as the average of 
$m$ and $m'$ in the centre of the lattice. $m_1$ and $m_2$ differ only because
of lattice effects. Hence, in the improved theory, the difference between
$m_1$ and $m_2$ is of order $a^2$.

\subsection{Aims}
We want to compute the critical quark mass at which the 
renormalized mass vanishes at
1--loop order of perturbation theory, which amounts to calculating 
$m_c^{(1)}$ in the series
\begin{equation}
m_c = m_c^{(0)} + m_c^{(1)}g_0^2 + {\rm O}(g_0^4).
\end{equation}
Since $m_1$ is only renormalized multiplicatively, 
it is sufficient to
require $m_1=0$.

Furthermore, we want to estimate the size of the lattice artefacts in the
mass calculation. In order to
do this, we compute two different discretization errors at 1--loop order of
perturbation theory. One is the difference
\begin{equation}
      d(L/a) = m_2(L/a) - m_1(L/a),
\end{equation}
the other one is the difference
\begin{equation}
e(L/a) = m_1(2L/a) - m_1(L/a).
\end{equation}

For these purposes, $f_A$ and $f_P$
have to be expanded up to 1--loop order.  
\section{$f_A$ AND $f_P$ AT 1--LOOP ORDER}
$f_A$ and $f_P$ may be written as
\beqn
&f_{A,P}(x_0) &= \cttilde^2\frac{a^9}{L^3}
\sum_{\mathbf{x},\mathbf{y},\mathbf{z}}\frac{1}{2}
\Bigl\langle \mbox{tr}\{P_+\Gamma P_- \nonumber\\
& &U(z-a\hat{0},0)S(z,x)\Gamma S(x,y)\nonumber\\
& &U(y-a\hat{0},0)^{-1}
\}\left. \Bigr\rangle_G\right|_{y_0=z_0=a}
\eeqn
where $\Gamma=\gamma_0\gamma_5$ for $f_A$ and $\Gamma=\gamma_5$
for $f_P$, and $S(x,y)$ is the quark propagator. $\cttilde$ is a coefficient
needed for ${\rm O}(a)$ improvement.\\
$S(x,y)$ and $U(x,\mu)$ are expanded up to order $g_0^2$, where $U(x,\mu)$
is expanded around the background field $V(x,\mu)$
\beqn
U(x,\mu) &=& V(x,\mu)\Bigl(1+g_0aq_{\mu}(x) \nonumber\\
& & +\frac{1}{2}g_0^2a^2q_{\mu}(x)^2
+ {\rm O}(g_0^3)\Bigr).
\eeqn
Collecting all contributions of order $g_0^2$ amounts to summing the following
diagrams~\cite{csw}:
  \noindent
  \begin{center}
  \begin{minipage}[b]{.3\linewidth}
     \centering\epsfig{figure=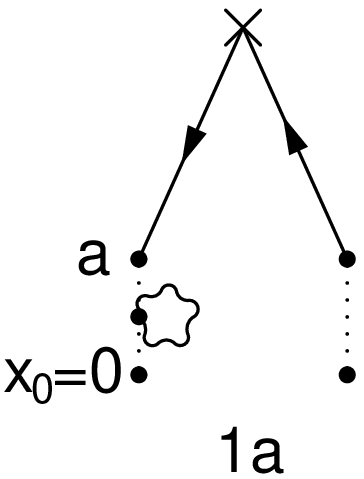,width=.8\linewidth}
  \end{minipage}
  \begin{minipage}[b]{.3\linewidth}
     \centering\epsfig{figure=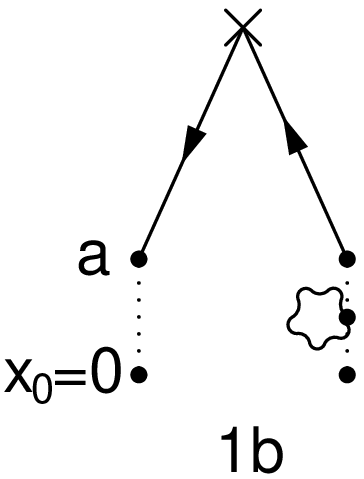,width=.8\linewidth}
  \end{minipage}
  \begin{minipage}[b]{.3\linewidth}
     \centering\epsfig{figure=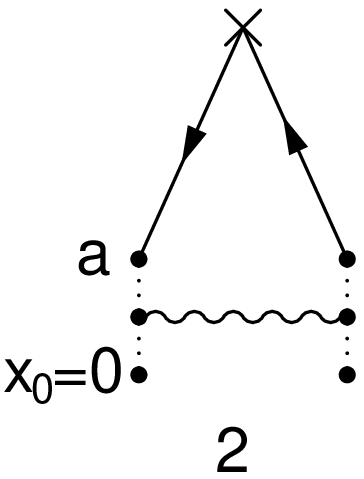,width=.8\linewidth}
  \end{minipage}\\
  \begin{minipage}[b]{.3\linewidth}
     \centering\epsfig{figure=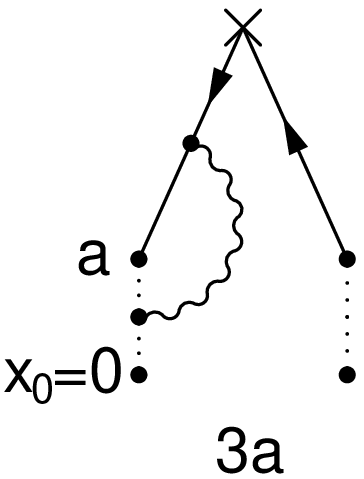,width=.8\linewidth}
  \end{minipage}
  \begin{minipage}[b]{.3\linewidth}
     \centering\epsfig{figure=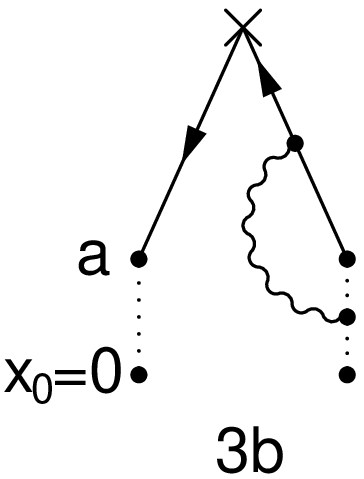,width=.8\linewidth}
  \end{minipage}
  \begin{minipage}[b]{.3\linewidth}
     \centering\epsfig{figure=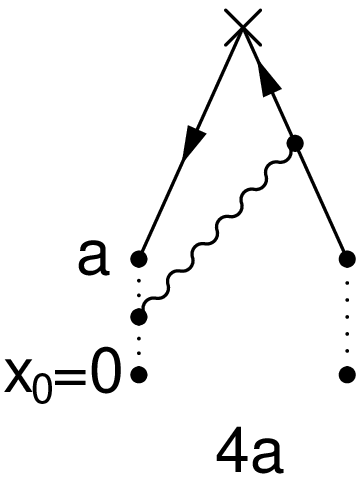,width=.8\linewidth}
  \end{minipage}\\
  \begin{minipage}[b]{.3\linewidth}
     \centering\epsfig{figure=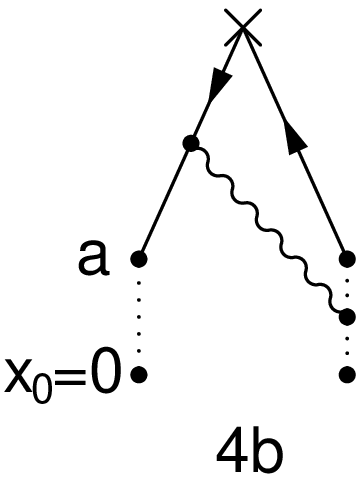,width=.8\linewidth}
  \end{minipage}
  \begin{minipage}[b]{.3\linewidth}
     \centering\epsfig{figure=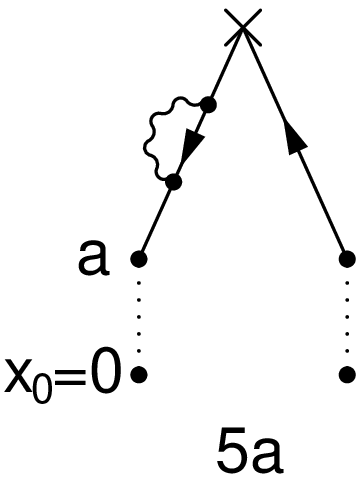,width=.8\linewidth}
  \end{minipage}
  \begin{minipage}[b]{.3\linewidth}
     \centering\epsfig{figure=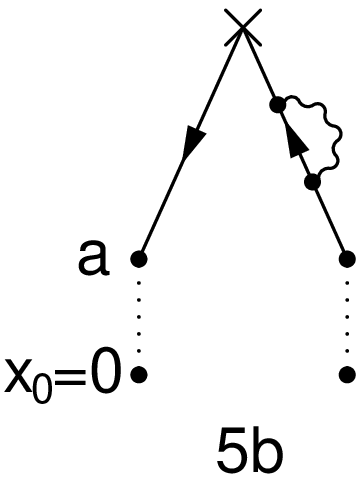,width=.8\linewidth}
  \end{minipage}\\
  \begin{minipage}[b]{.3\linewidth}
     \centering\epsfig{figure=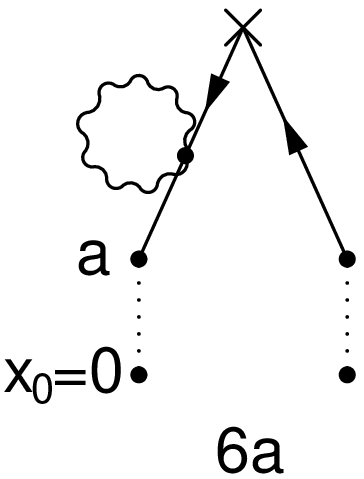,width=.8\linewidth}
  \end{minipage}
  \begin{minipage}[b]{.3\linewidth}
     \centering\epsfig{figure=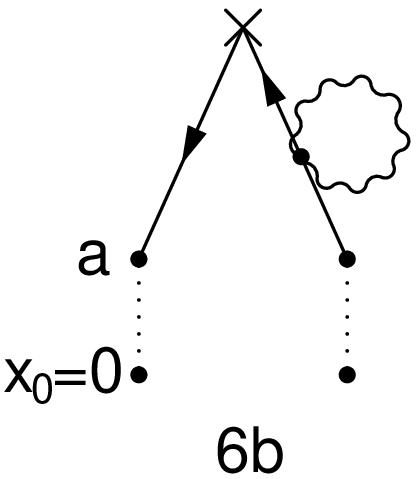,width=.8\linewidth}
  \end{minipage}
  \begin{minipage}[b]{.3\linewidth}
     \centering\epsfig{figure=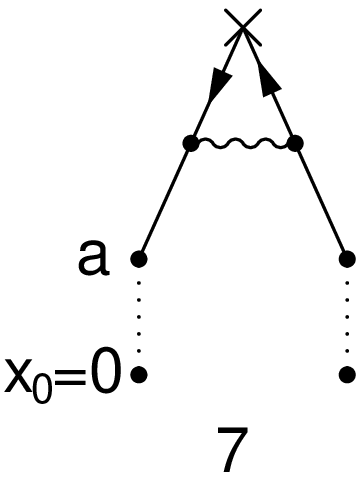,width=.8\linewidth}
  \end{minipage}\\
  \begin{minipage}[b]{.3\linewidth}
     \centering\epsfig{figure=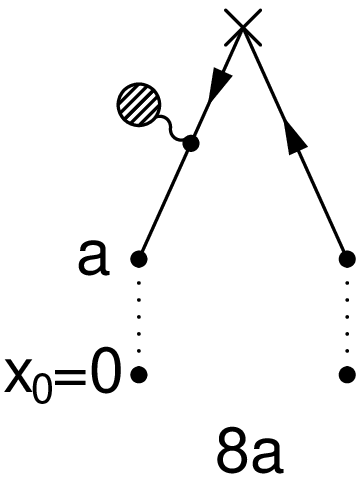,width=.9\linewidth}
  \end{minipage}
  \begin{minipage}[b]{.3\linewidth}
     \centering\epsfig{figure=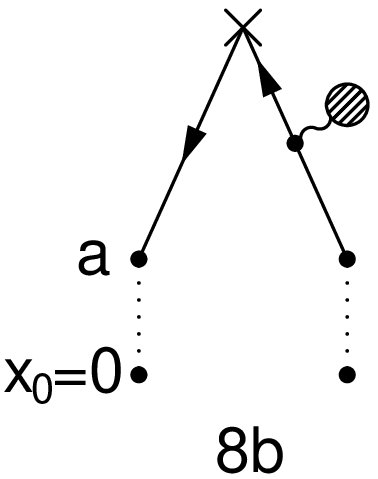,width=.9\linewidth}
  \end{minipage}\\
  \begin{minipage}[b]{.3\linewidth}
     \centering\epsfig{figure=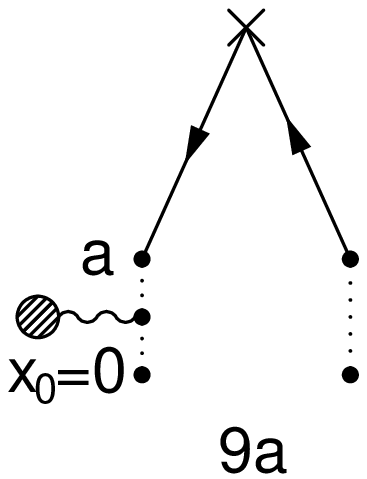,width=1.15\linewidth}
  \end{minipage}\hspace{4mm}
  \begin{minipage}[b]{.3\linewidth}
     \centering\epsfig{figure=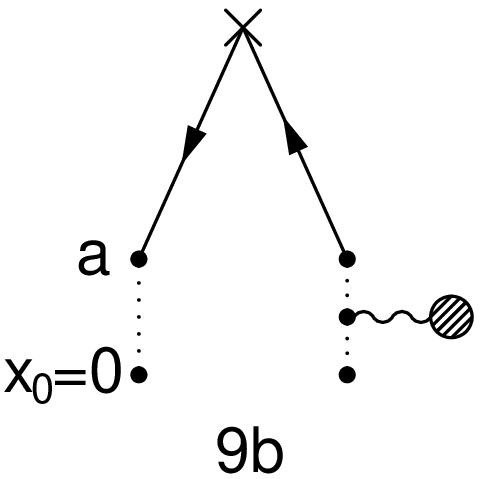,width=1.15\linewidth}
  \end{minipage}\\
  \vspace{4mm}
  \begin{minipage}[b]{.5\linewidth}
     \centering\epsfig{figure=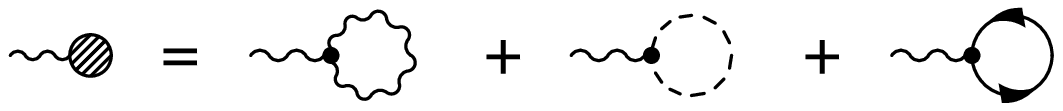,width=1.1\linewidth}
  \end{minipage}
  \end{center}
The dotted lines are the links between $x_0=0$ and $x_0=a$, and the cross
denotes the insertion of the axial current or density.
\section{RESULTS}
\subsection{The critical quark mass}
The results for the critical mass at 1--loop order are shown below for the
case of two light flavours, the quenched case and the bermion case
(i.e.~$N_{\rm f}=-2$)~\cite{Nfm2}. They seem to converge quickly to the $N_{\rm f}$ independent
continuum limit $am_c^{(1)} = -0.2700753495(2)$~\cite{wohlert,axial}.
%\begin{figure}
%  \noindent
  \begin{center}
    \epsfig{figure=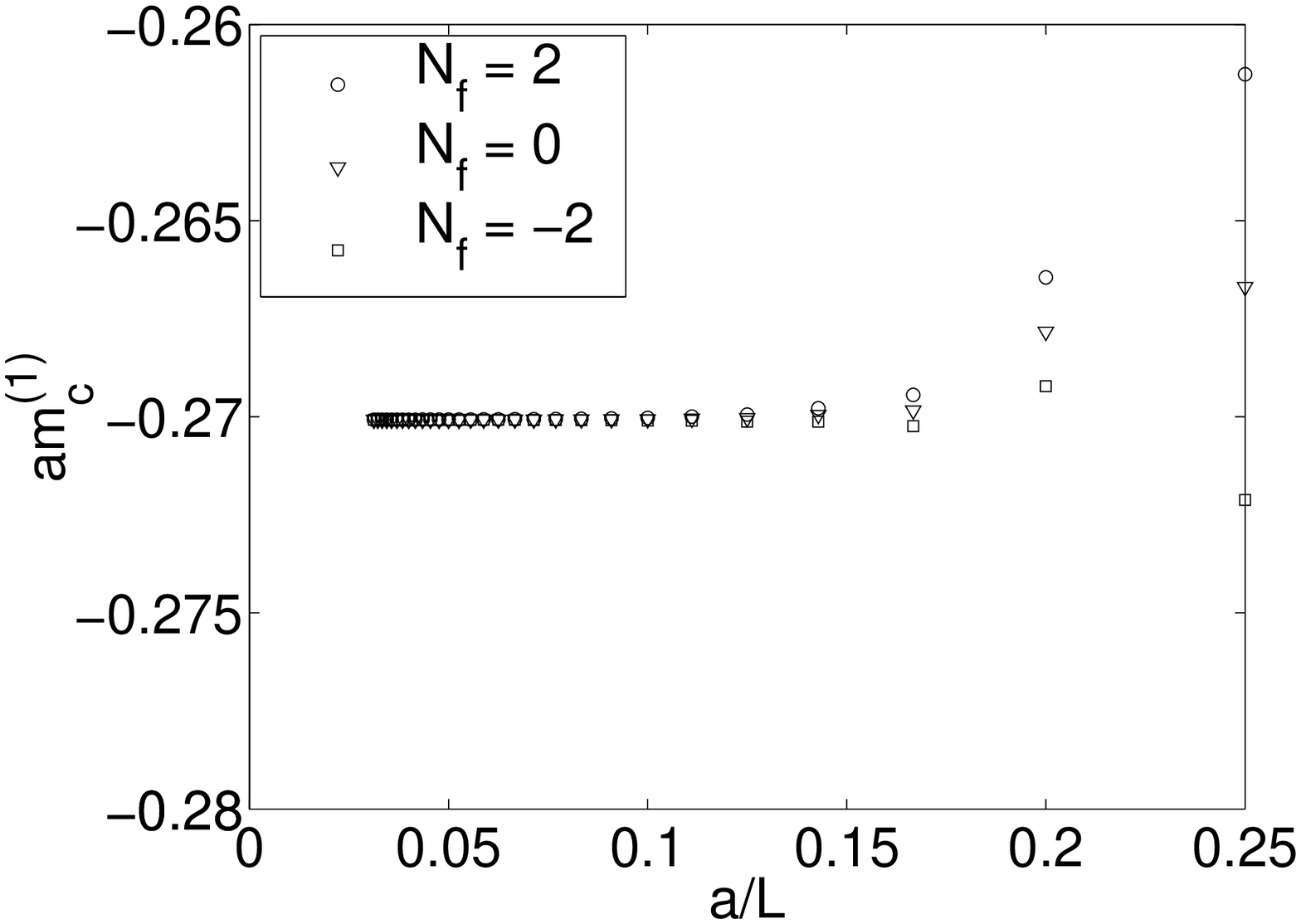,width=\linewidth}
  \end{center}
%  \caption{The critical quark mass at 1--loop order at different flavour numbers}
%  \label{fig:mcrit}
%\end{figure}
%
%\begin{figure}[htbp]
%  \noindent
\subsection{Lattice artefacts}
The results for the lattice artefacts are shown below.

As one expects for discretization errors, the continuum limit is zero.
  The lattice artefacts are small at 1--loop level, but the perturbative results
  for $m_2(L/a)-m_1(L/a)$ appear to be bigger than the results obtained by
  Monte Carlo simulations~\cite{Nf2}.
  \begin{center}
    \epsfig{figure=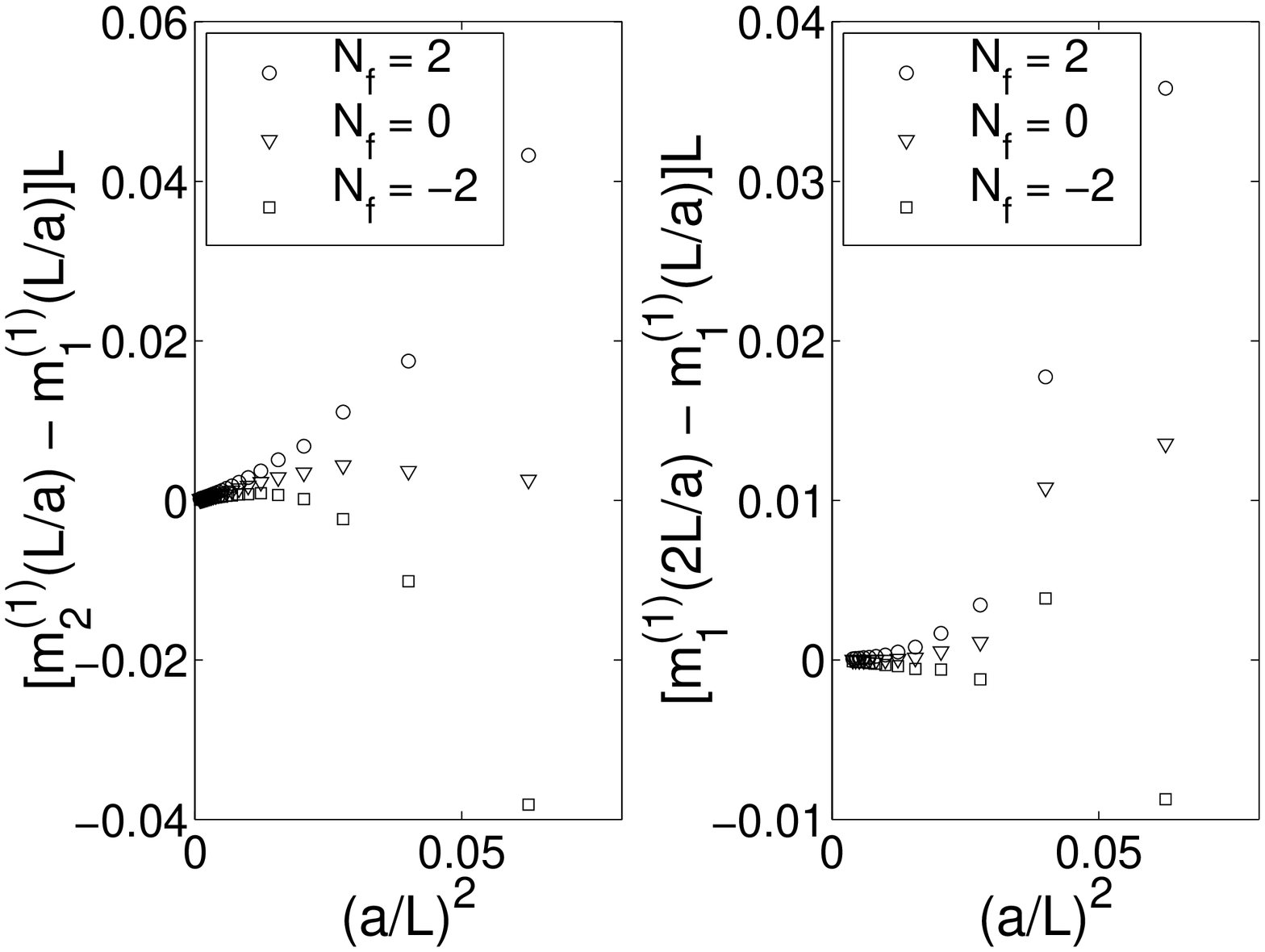,width=\linewidth}
  \end{center}
%  \caption{Lattice artefacts at 1--loop order}
%  \label{fig:art}
%\end{figure}
%
\section*{ACKNOWLEDGEMENTS}
We would like to thank Peter Weisz for essential checks on the calculations.
This work is supported by the Deutsche Forschungsgemeinschaft under
Graduiertenkolleg GK271.
\end{document}